\documentclass[10pt,final,conference]{IEEEtran}
 \usepackage{eso-pic}
 \usepackage{color}
 \definecolor{DisclaimerGray}{gray}{0.92}
 
 \AddToShipoutPicture*{\put(45,760){\colorbox{DisclaimerGray}{\parbox{\textwidth}
 {\footnotesize \copyright IEEE, 2018. This is the author's version of the work. Personal use of this material is permitted. However, permission to reprint/republish this material for advertising or promotional purpose or for creating new collective works for resale or redistribution to servers or lists, or to reuse any copyrighted component of this work in other works must be obtained from the copyright holder. The definite version is published at IEEE ENERGYCON 2018.}}}}
 
\usepackage{hyperref}
\hypersetup{
   pdfauthor={ {Midhat Jdeed, Ekanki Sharma, Christoph Klemenjak, Wilfried Elmenreich} },
   pdfkeywords={Smart Grid, Grid Modelling, RAPSim, GridLAB-D, Power Flow},
   pdftitle={Smart Grid Modeling and Simulation -
Comparing GridLAB-D and RAPSim via two Case Studies}
 }

\usepackage[english]{babel}
\usepackage[utf8x]{inputenc}
\usepackage[T1]{fontenc}

\usepackage{siunitx}
\usepackage{amsmath}
\usepackage{graphicx}
\usepackage[colorinlistoftodos]{todonotes}

\begin{document}

\title{Smart Grid Modeling and Simulation -
Comparing GridLAB-D and RAPSim via two Case Studies
}


\author{
Midhat~Jdeed, Ekanki~Sharma, Christoph~Klemenjak, and~Wilfried~Elmenreich\\
\IEEEauthorblockA{Institute of Networked and Embedded Systems/Lakeside Labs\\
Alpen-Adria-Universit\"at Klagenfurt\\
9020 Klagenfurt, Austria\\
\{\emph{name.surname}\}@aau.at}

}

\IEEEoverridecommandlockouts
\IEEEpubid{\makebox[\columnwidth]{978-1-5386-3669-5/18/\$31.00~
\copyright2018
IEEE \hfill} \hspace{\columnsep}\makebox[\columnwidth]{ }} 

\maketitle
\begin{abstract}
One of the most important tools for the development of the smart grid is simulation. Therefore, analyzing, designing, modeling, and simulating the smart grid will allow to explore future scenarios and support decision making for the grid's development. In this paper, we compare two open source simulation tools for the smart grid, GridLAB-Distribution (GridLAB-D) and Renewable Alternative Power systems Simulation (RAPSim). The comparison is based on the implementation of two case studies related to a power flow problem and the integration of renewable energy resources to the grid. Results show that even for very simple case studies, specific properties such as weather simulation or load modeling are influencing the results in a way that they are not reproducible with a different simulator.

\end{abstract}

\section{Introduction} \label{sec:introduction}

An increasing demand for energy generation from alternative renewable sources, integrating energy storage, and the need for increasing energy efficiency via management systems are key motivations for transforming the conventional grid to a smart grid. The concept of the smart grid has been developed to make power systems more environmental friendly by using renewable energy sources, more reliable even during disasters or sudden faults by enabling consumers to supply energy. During this transformation, major challenges must be overcome. For instance, integrating a variety of renewable energy resources such as wind turbines, photovoltaic systems with the power grid causes instability in the grid. In addition, integrating information and communication technologies result in new challenges such as the need to provide dynamic pricing in real time, security issues, and non-intrusive load monitoring via smart metering \cite{pochacker2013simulating}. To create a functional smart grid, it is necessary to predict the behavior of the energy grid when certain parameters are changed. Simulation is fundamental to fulfill this requirement. A smart grid simulator must simulate two-way communication between the utility and the consumers to study a wide range of planning and operational situations such as power generation and transmission expansion planning.
Smart grid simulation is an essential prerequisite for evaluation and analysis before establishing a real-world smart grid. This can be used to optimize the overall performance of the grid, specifically, when integrating renewable energy sources. For example, one important application of solar energy is in Photovoltaic Water Pumping Systems (PVPS), which are required in rural areas. Many economic benefits can be garnered by developing a model which optimizes the size and performance of the solar panels by considering two basic factors: meteorological data and the average hourly water flow rate at the intended site \cite{muhsen2016multiobjective}.

Before the smart grid, power grid simulations were mostly concerned with estimating power losses and end user load for a given power flow. The transition toward the smart grid has increased the level of complexity of the energy system, because of the integration of distributed and renewable energy sources, smart meters, smart appliances, electric vehicles, etc. into the electric grid \cite{pochacker2013simulating}. Therefore, new tasks are required from the simulator such as the optimization of distributed energy resource management when the smart grid is in operation \cite{ramachandran2011intelligent}, stand-alone PV systems~\cite{kathib-elmenreich:16}, demand response in deregulated electricity markets \cite{albadi2008summary}, smart meters \cite{depuru2011smart}, energy storage in electric vehicles using smart grid \cite{van2014increasing}, predicting consumer demand \cite{lu2010smartgridlab,frincu2014accurate}, and integrating power and communication networks \cite{mets2014combining,razaq2015simulating,bian2015real,hansen2016enabling}. 

In this paper, two open-source smart grid simulation tools are compared, GridLAB-D and RAPSim. This paper provides a short description of these simulation software tools along with their functions. Two case studies are presented as a means of comparison between their respective performances.

The paper is structured as follows:
Section \ref{sec:related} presents related work. Section \ref{sec:modeling} introduces relevant aspects in the domain of modeling and simulation  power flow. In Section \ref{sec:cases}, we present the case studies in the field of renewable energy generation and power distribution. A discussion about the two simulators, GridLAB-D and RAPSim, is introduced in \ref{sec:dis}. Finally, Section \ref{sec:conclusion} concludes the paper and provides an outlook on future work.

\section{Related Work} \label{sec:related}


Software for smart grid simulation can be divided into two main categories: 
The first is \emph{commercial} software which is expensive, highly specialized and difficult to modify, which is thus less suitable for research and teaching. The second is \emph{free} software, which is usually available as an open source, its easy to modify according to the user's requirements, hence it is very suitable for research and teaching purposes. There are several simulators which are available as free of cost. In \cite{pochacker2013simulating}, ten free simulators such as GridLAB-D, AMES, InterPSS, OpenDSS, MatPower etc. are compared and a test case is made for the four selected simulators (GridLAB-D, MatPower, PSAT, and InterPSS) using the same model IEEE 14-bus system to compare the results and the methods employed by the simulators.

In the past years, several smart grid software tools addressing power flow problems and integration of renewables into the existing power grid have been published. The IEEE has published reference models for simulation purposes, which have been used by researchers in different simulations. The authors of \cite{ahourai2013grid} implemented an electric vehicle (EV) load within GridLAB-D using the IEEE 13 Node Test Feeder to analyze the impact of  EV charging under various penetration rates\footnote{\url{http://sites.ieee.org/pes-testfeeders/resources/}}. \cite{owuor2011ieee} uses the IEEE 34 node radial test feeder as a simulation test bench to study the impact of distributed generation on a distribution system using DigSILENT PowerFactory. 
\cite{moffet2011review} reports a good review on open source power grid simulation tools which includes mainly three smart grid simulation software, namely GridLAB-D, Open Distribution System Simulator (OpenDSS) and Parametric Analysis of Power Grids with Matlab (APREM). Two case studies are explored and the respective performances of the software tools are compared. 
 
The authors of \cite{pochacker2014microgrid} introduce a novel simulator for smart grid simulation, RAPSim. The open-source microgrid simulation framework RAPSim aims to provide a better understanding of power flow behaviour in smart microgrids. In particular, the integration of renewable energy sources plays an important role. RAPSim allows simulating grid-connected or standalone microgrids with solar, wind or other renewable energy sources. The framework computes the power generated by the individual sources in the microgrid and is able to conduct a power flow analysis. Therefore, RAPSim allows determining the optimal placement of distributed generation units in a microgrid.

To the best of our knowledge, there is no comparative study between GridLAB-D and RAPSim. This paper aims to close this research gap and compare these two simulation tools via two case studies.

\section{Power Flow Simulation}  \label{sec:modeling}

Simulation of the power flow is a basic tool for analyzing, operating and planning of the smart grid (as it was for transmission power systems before).
In general, there are four basic quantities in a power system:
\begin{enumerate}
\item The magnitude of the voltage |V|
\item The voltage angle $\theta$
\item The real power injection P
\item The reactive power injection Q
\end{enumerate}
To study and analyze the power flow, two of these above-mentioned quantities are provided and the remaining two are obtained by forming a set of nonlinear algebraic equations. Then, by choosing a suitable mathematical method (e.g., the Gauss-Seidel method, the Newton-Raphson method, fast-decoupled load-flow (FDLF), or the forward-backward algorithm) is applied to solve the power flow problem iteratively. Most simulators use Newton-Raphson method or the Gauss-Seidel method.

\section{Case Studies} \label{sec:cases}
\begin{figure} [ht]
 \begin{center}
 		\centerline{\includegraphics[width=\columnwidth] {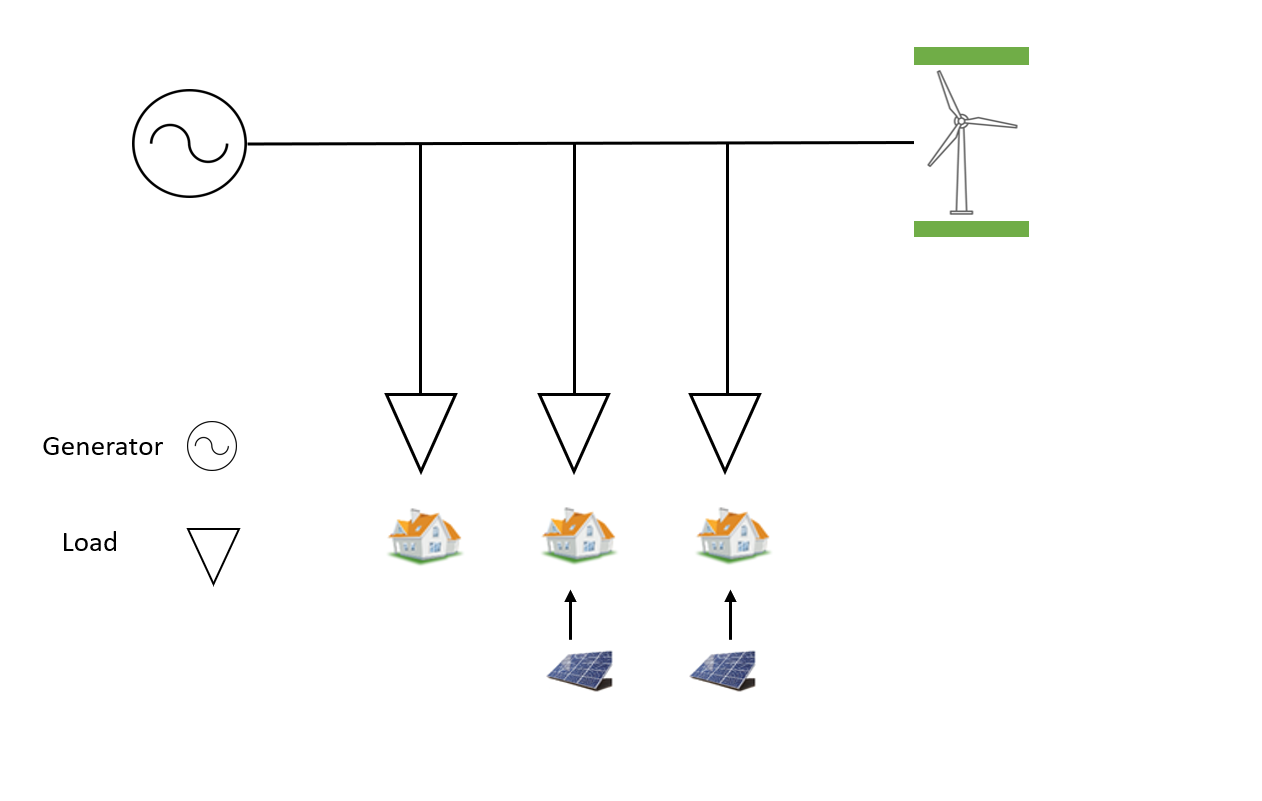}}		
        \vspace{-1em}
		\caption{The first case-study: Renewable Energy Generation.}
		\label{Fig:1}
        \vspace{-1em}
 \end{center}

\end{figure}

The performance of two open-source smart grid simulation tools RAPSim and GridLAB-D is compared by implementing two case studies, which focus on calculating the power flow in a microgrid and a distribution grid example. The former case includes a scenario where a microgrid is connected to grid, hence working in grid-connected mode. In this mode, microgrid can sell off or buy the excess energy to or from the utility grid \cite{sobe2013smart}. The latter case includes a modified scenario where a PV system has been added to 
explore the effect of local generation on voltage levels.
In both cases, a period of two days has been simulated and observed at intervals of one hour.   
Fig. \ref{Fig:1} illustrates the first case study and shows the placement of load and decentralized power generation in a microgrid.  
The aim of the first case study is to assess the power distribution among the power generators, the impact of adding renewable energy sources and to calculate the amount of available power production and the amount of consumption, considering production changes due to weather information (such as wind speed and cloud coverage). The power generators considered for this case study are a small wind turbine with a peak power production \SI{1500}{\watt} and two small solar panels with a peak power production of \SI{500}{\watt}. This case study does not consider power losses along the distribution lines. However, the first case study demonstrates how the power production gets distributed among the different generators, especially during the night when no power is produced by the solar panels. 

\begin{figure} [ht]
 \begin{center}
 		\centerline{\includegraphics[width=\columnwidth] {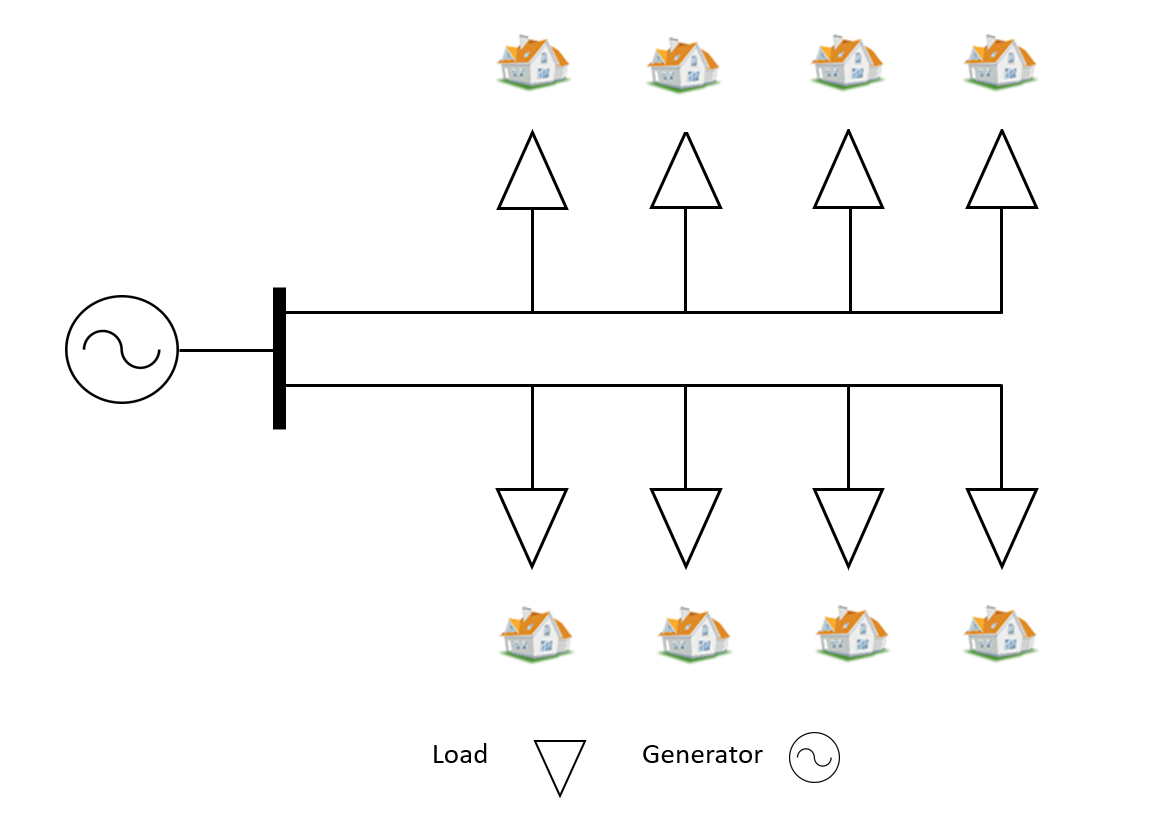}}		
		\caption{The second case-study: Power Distribution in a Street}
		\label{Fig:2}
        \vspace{-2em}
 \end{center}
\end{figure}

The second case study as shown in Fig. \ref{Fig:2}, consists of a feeder connected to two radial powerlines connected to houses.
The aim of this case study is to simulate the voltage distribution before and after adding solar panels. 
The main parameters for this case study are as follows:
\begin{itemize}
\item The nominal voltage at the feeder is \SI{230}{\volt} for both simulators.
\item Single phase distribution lines with a distance of \SI{150}{\meter} for each street are considered. The distance is \SI{60}{\meter} until the connection point for the first house and then \SI{30}{\meter} each between the following connection points. Each house is connected with a tap line of \SI{10}{\meter}.
\item  The case study used copper as a conductor where its resistivity is $1.724 \cdot 10^{-8}$\,$\Omega$m at temperatures of \SI{20}{\celsius}. The cross-sectional  is 150 mm$^{2}$, which corresponds to a typical feed cable in a street \cite{elmenreich2013demand}. Therefore, we get a resistance value 0.115 $\Omega$/km or 0.185 $\Omega$/mile. Line inductances were not included in the model. 
\item Eight households with demands between \SI{4}{\kW} and \SI{6}{\kW}.
\end{itemize}

\subsection{Simulation with RAPSim} \label{RAPSim}

The open-source microgrid simulation tool RAPSim \cite{pochacker2014microgrid} has been developed at the Institute of Networked and Embedded Systems of the Alpen-Adria-Universität Klagenfurt. The software features a user-friendly graphical interface that allows to set up scenarios using a drag and drop approach. This tool aims at simulating microgrids with renewable energy sources. In this paper, RAPSim version V0.95 is used. The main characteristics of RAPSim are:
\begin{itemize}
\item It is designed to simulate and analyze power flow in smart microgrids in both on-grid and off-grid mode. Therefore, RAPSim is helpful to optimize power flow and improve distributed generation units present in the grid.
\item RAPSim has a graphical user interface for creating, saving the scenarios and loading them in a XML format. Furthermore, simulation results can be saved by generating an output file as a CSV format which allows further processing by other tools.
\item RAPSim allows to chose from several power flow algorithms such as Simple Power Distribution, ACPowerFlow (AC-PF). 
\item Since RAPSim is free and open source, it has huge potential for developing new models and algorithms, as well as modifying existing ones. For instance, \cite{pochacker2015model} presented an implementation for a new customized wind turbine model with three short code snippets.
\end{itemize}

\subsubsection{The first case study in RAPSim}
\begin{figure} [ht]
 \begin{center}
		{\includegraphics[width = 0.7\columnwidth] {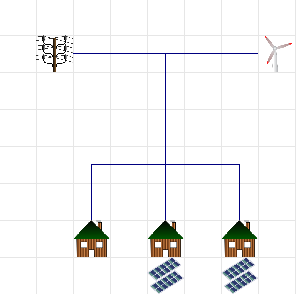}}		
		\caption{The first case modeled in RAPSim}
		\label{Fig:3}
        \end{center}
\end{figure}

\begin{figure} [ht]
 \begin{center}
 		\includegraphics[width = 1\columnwidth] {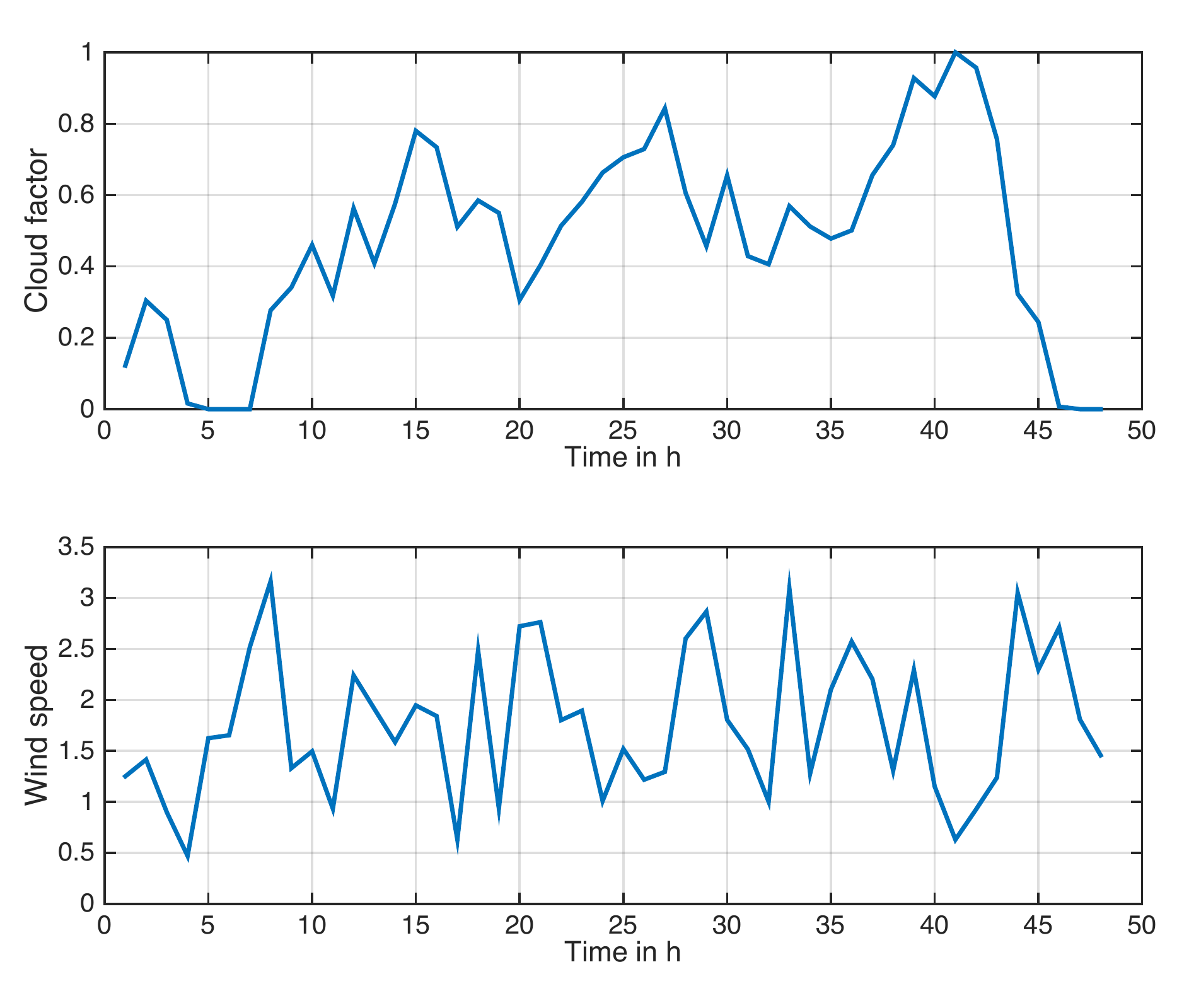}	
        \vspace{-2em}
		\caption{Cloud factor and wind speed}
		\label{Fig:4}
        \vspace{-0.5em}
 \end{center}
 
\end{figure}

In the first case study, the following objects have been added: Three houses with constant load demand, a substation, two solar panels and a wind turbine. These are placed in the lattice in RAPSim as shown in Fig. \ref{Fig:3}.
The power produced by renewable energy sources especially the wind and solar energy sources depend highly on the weather conditions. So meaningful weather input is very important for simulation. The weather class in RAPSim is using a stochastic model for generating and delivers three parameters cloudiness, wind speed and temperature.  Wind speed is modeled using Weibull probability density function and the cloud factor is modeled as random toy model. The cloud coverage and wind speed affects the output of solar panels and wind turbine, respectively, so its important to consider them while modeling the scenarios.
A simple power distribution algorithm has been chosen for the first case study. We simulating two days with a resolution time of one hour. Fig. \ref{Fig:4} illustrates the cloud factor and the wind speed values with respect to the time steps (from 1 to 48 hours).

\begin{figure} [ht]
 \begin{center}
		\centerline{\includegraphics[width = 1\columnwidth] {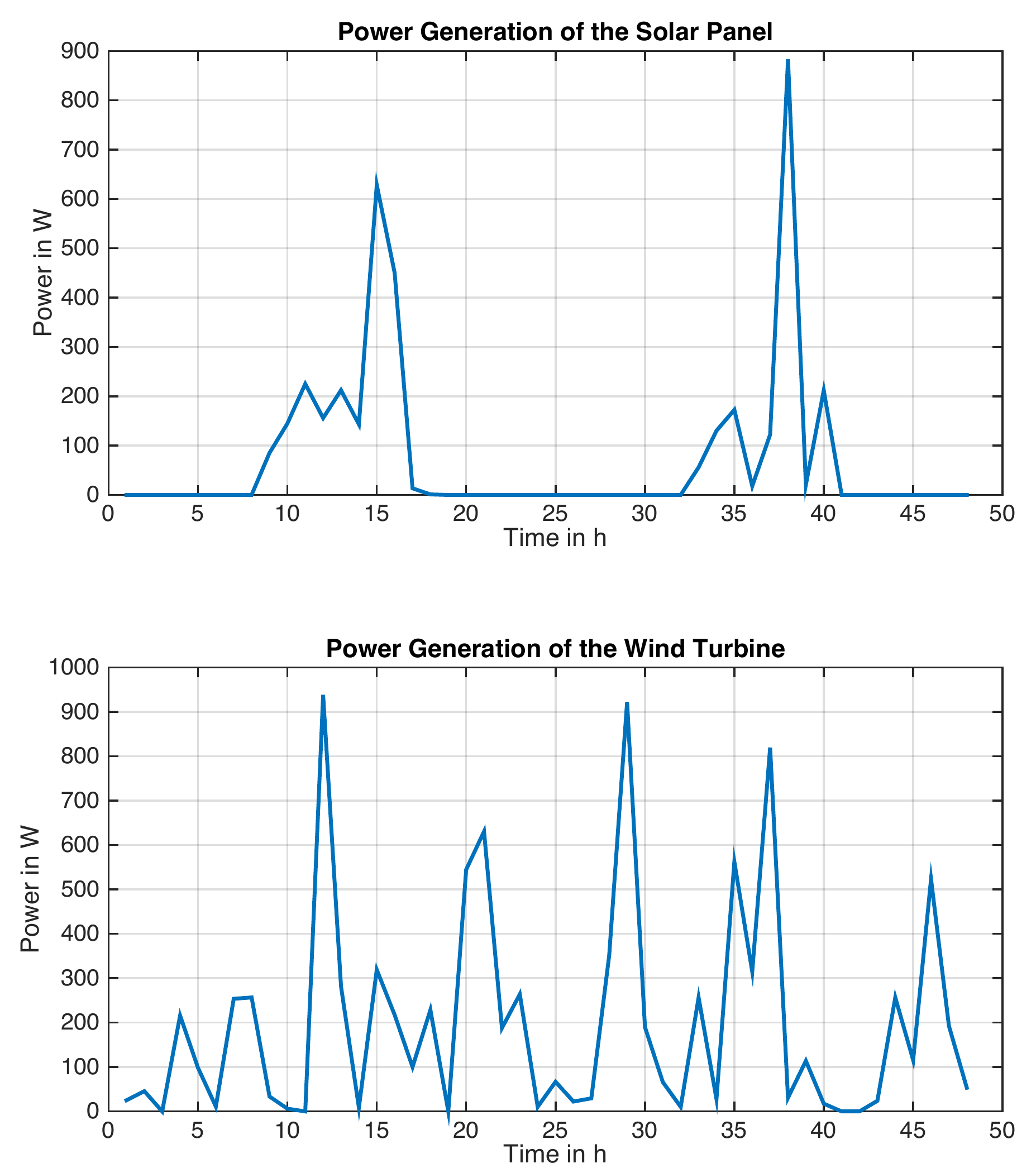}}		
		\caption{RAPSim: Power production}
		\label{Fig:5}
         \end{center}
         \vspace{-0.5em}
\end{figure}

Fig. \ref{Fig:5} shows power production from both the generators for two days. The graph shows high correlation between cloud coverage and solar output power and between wind speed and wind power output. During early morning hours and night hours the intensity of sunlight is minimum as compared to the daytime hence solar panel doesn't produce any output. Whereas, wind turbines produce power as they depend on the wind speed. 

\subsubsection{The second case study in RAPSim}
\begin{figure*} [ht]
		\centerline{\includegraphics[width=.88\textwidth] {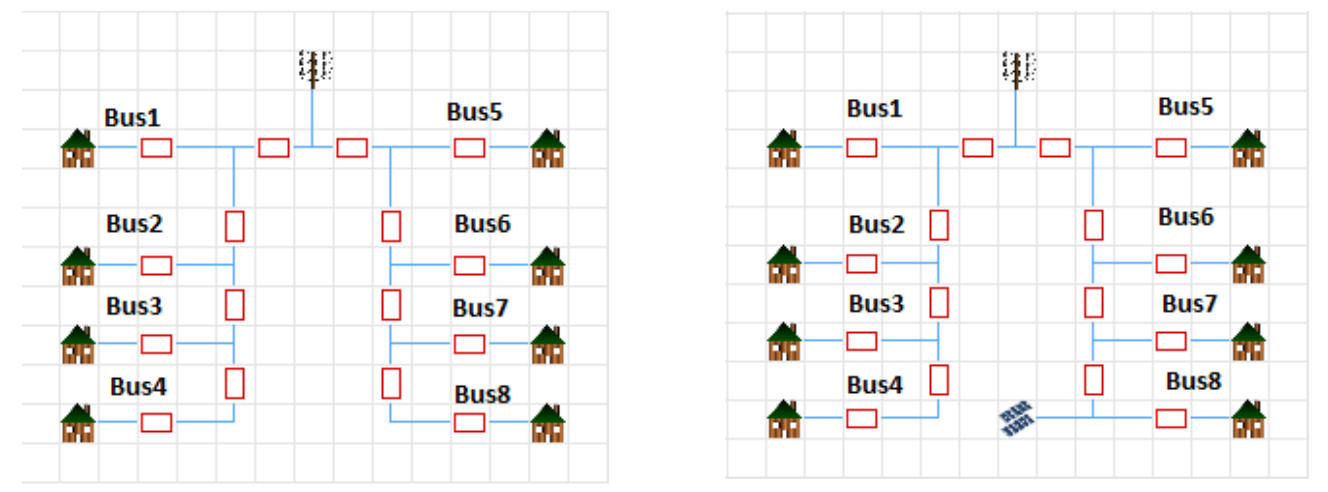}}		
		\caption{The second case modelled in RAPSim}
		\label{Fig:6}
        \vspace{-0.5em}
\end{figure*}

\begin{figure} [t]
\centerline{\includegraphics[width = .85\columnwidth] {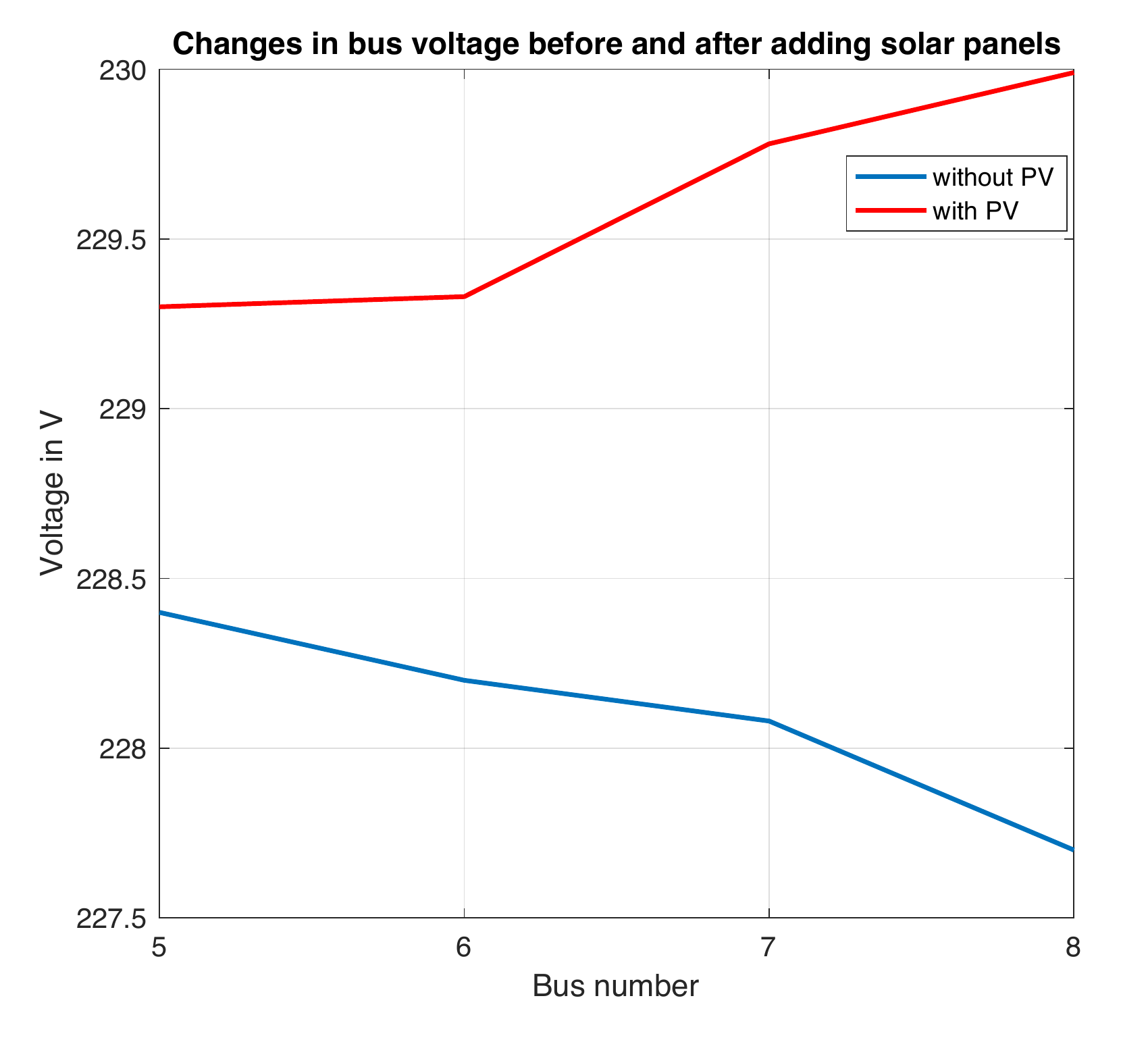}}		
	\vspace{-0.5em}
		\caption{RAPSim: Voltage drop across the bus}
		\label{Fig:7}
        \vspace{-1em}
\end{figure}

The idea of this case study is to analyze the change in the voltage before and after adding the solar panels. To simulate the same in the first part, houses with constant load demand are added to the feeder. In the second part a solar panel has been added to Bus 8 (Fig. \ref{Fig:6}). In this case study, the AC-PF solver has been used. 
Fig. \ref{Fig:7} illustrates the change in the voltage over a period of two days considering the added solar panel in the scenario. In the first scenario, where only consumers are connected to the feeder, bus 8 has the minimum voltage. In the second scenario, the PV panel causes an increase of the voltage nearby the PV.  

\subsection{Simulation Using GridLAB-D} \label{GridLAB-d} 

GridLAB-D is an open source modeling and simulation tool developed by the United States Department of Energy that integrates detailed power systems and end-use models \cite{schneider2009distribution}. Furthermore, GridLAB-D can simulate distributed power networks and analyze the impact of integrating renewable energy sources.
GridLAB-D is an agent-based simulator and capable of independently modeling the changes of devices or agents along the grid. It also models and simulates the interaction between different agents \cite{chassin2014gridlab}.
However, to simulate a model in GridLAB-D, the simulation setup must be described in a .glm file. This file is then used as an input in GridLAB-D. GridLAB-D is operated via a console. There is no graphical user interface which makes working with GridLAB-D more difficult as compared to RAPSim which has a more user-friendly interface. 
The main characteristics of GridLAB-D\footnote{\url{http://www.gridlabd.org/}} are:
\begin{itemize}
\item GridLAB-D contains modules with six packages of classes. Depending
on these packages, the related modules must be imported for the scenario that
is being simulated. For example, to import objects related to renewable energy sources, the generator module must be imported.
\item GridLAB-D offers time-series simulation with resolution from seconds through decades.
\item GridLAB-D works with third-party data management and analysis tools. 
\end{itemize}
The weather data file (WA-Yakima.tmy2) which is used for this case-study is available and can be downloaded from GitHub repository of GriDLAB-D\footnote{\url{https://raw.githubusercontent.com/gridlab-d/gridlab-d/master/models/WA-Yakima.tmy2}}. TMY is an acronym for typical meteorological year. In this file, weather data for a particular location is averaged to provide a typical baseline for the weather of a particular geographical location on a given day at a given hour \cite{tenney2008gridlab}.

\subsubsection{The first case study in GridLAB-D}
\begin{figure} [t]
 \begin{center}

		\centerline{\includegraphics[width = 1\columnwidth] {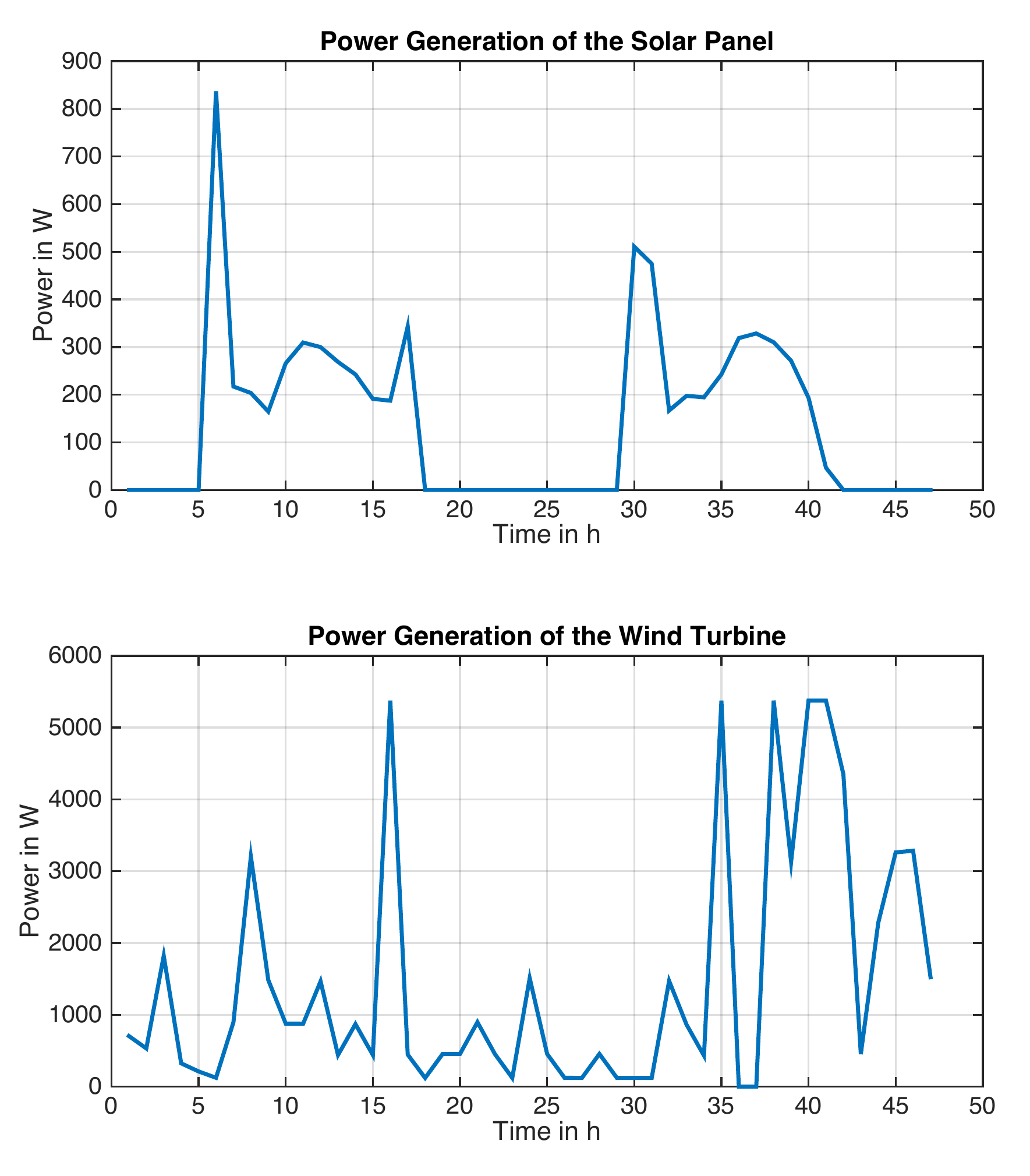}}
        \vspace{-.3em}
		\caption{GridLAB-D: Power production}
		\label{Fig:8}
        \vspace{-3.5em}
         \end{center}
\end{figure}

For defining the constant value for load demand, a triplex-load object has to be used instead of a house object. Furthermore, some of the objects cannot be placed directly in GridLAB-D as compared to RAPSim. For instance, for placing the solar panels in GridLAB-D, an inverter object has to be specified as well, as it is not possible to connect solar panels directly to the triplex meters. Moreover, all the DC power generating resources should have an inverter as its parent object.  
Fig. \ref{Fig:8} illustrates measurements for a simulation period of two days indicating power produced by solar panels and wind turbine. 
\begin{figure} [t]
 \begin{center}

 		\centerline{\includegraphics[width = \linewidth] {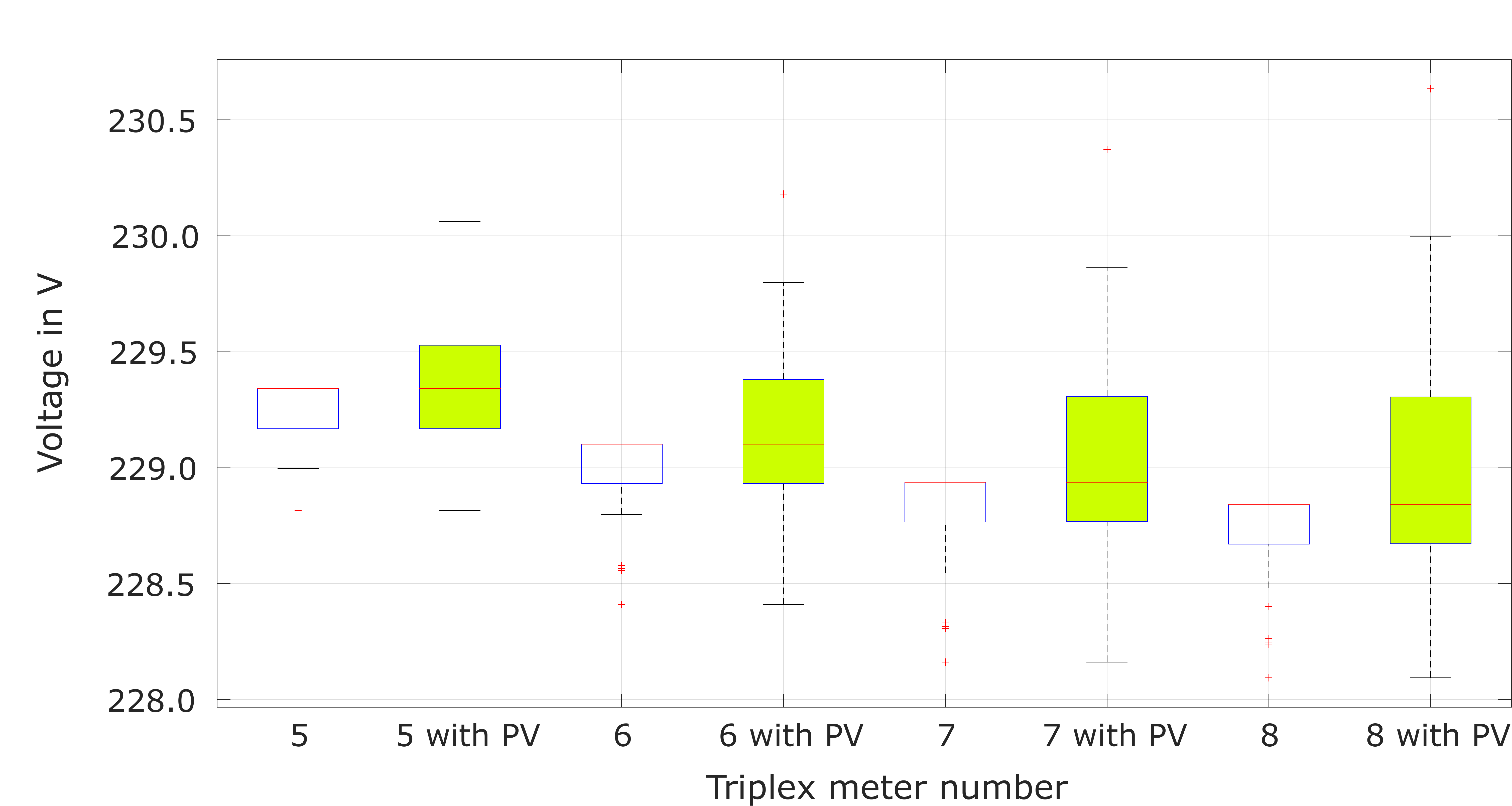}}		
		\caption{GridLab-D: Voltage drop across the triplex meters}
		\label{Fig:9}
        \vspace{-2em}
         \end{center}
\end{figure}
\subsubsection{The second case study in GridLAB-D}
In the second case study, the resistance is considered for the transmission lines and solar panels are added to one of these two lines to determine the impact on voltage on the triplex meter. Fig. \ref{Fig:9} illustrates the measured voltage for simulation period of 48 hours without considering solar panels in first case and the impact on measured voltage after adding solar panels.

\section{Discussion} \label{sec:dis}
GridLAB-D has been used more widely than RAPSim. RAPSim is designed for microgrid simulation, whereas GridLAB-D is more oriented towards the whole smart grid. RAPSim provides lattice visualization with a user-friendly interface to place eight types of objects: solar panels, wind turbines, houses, power lines, connectors, fuel generators, power plants, and grid connections. The properties for these objects are extensible which holds great promise for developers. GridLAB-D offers many specific objects for modeling power flow such as capacitors, regulators, inverters and meters. However, GridLAB-D has no GUI which makes implementing scenarios more difficult and requires greater expertise from users.

For the first case study, both simulators were used to simulate power distributed generation as a simple calculation with a time resolution of one hour. Results in Figure~\ref{Fig:5} and Figure~\ref{Fig:8} show similar patterns, but due to using different weather the results are not identical. 

For the second case study, both simulators were used to model the impact of voltage of adding a solar panel in the scenario. The configuration for the second case study in GridLAB-D was more complex than in RAPSim, since it required a lot of hand-written code defining nodes and powerlines. The properties of a house object in GridLAB-D have no parameter to specify a constant load as it is possible in RAPSim. The demand for a house in GridLAB-D model is thus fluctuating over time. Adding an object with a constant load is possible in GridLAB-D, we can either use a triplex-load object connected to the power flow or use a zipload object and specify a constant power load. However in that case, studying the impact of adding a solar panel was not possible. To our knowledge, adding solar panels with its inverter could only be attached to a triplex-meter or meter object and any other objects like triplex-load, solar panels work in "stand-alone" mode, where it doesn't interface with the powerflow model. As a consequence, the results of case study 2 as shown in Figures~\ref{Fig:7} and \ref{Fig:9} cannot be easily compared. To indicate the fluctuation of the values, we have used a boxplot representation for the GridLAB-D as depicted in Figure~\ref{Fig:9}.

\section{Conclusion} \label{sec:conclusion}
This paper has shown how two simple case studies can be modeled and simulated in two open-source simulators, GridLAB-D and RAPSim. The simulators have a number of similarities, including being free and open source, as well as offering power simulation and the ability to simulate distributed and renewable energy sources. GridLAB-D is more suited to studies that include modeling and simulating residential loads and power flow optimization. RAPSim is suited to the classroom environment and teaching the main concepts for the smart grid and simulating the microgrid. Some modeling components, e.g. the weather module or the load behavior of a typical house are different in the two simulators, so that the same scenario yields different results. For instance, in the case of GridLAB-D, adding a PV system in the same way as in RAPSim was not possible, since GridLAB-D enforced a more detailed, realistic model, which could on the one hand be seen as a feature or on the other hand is the enforcement of complex models a problem for reproducibility across different simulators. In the future we expect to have test cases emerging which can be used on a variety of simulators yielding the same results, similar to the IEEE node test cases for power flow simulation.

\section*{Acknowledgments}

We would like to thank the anonymous reviewers for their careful reading of our manuscript and their  
insightful comments and suggestions.

\bibliographystyle{IEEEtran}
\bibliography{sample}

\begin{thebibliography}{10}
\providecommand{\url}[1]{#1}
\csname url@samestyle\endcsname
\providecommand{\newblock}{\relax}
\providecommand{\bibinfo}[2]{#2}
\providecommand{\BIBentrySTDinterwordspacing}{\spaceskip=0pt\relax}
\providecommand{\BIBentryALTinterwordstretchfactor}{4}
\providecommand{\BIBentryALTinterwordspacing}{\spaceskip=\fontdimen2\font plus
\BIBentryALTinterwordstretchfactor\fontdimen3\font minus
  \fontdimen4\font\relax}
\providecommand{\BIBforeignlanguage}[2]{{%
\expandafter\ifx\csname l@#1\endcsname\relax
\typeout{** WARNING: IEEEtran.bst: No hyphenation pattern has been}%
\typeout{** loaded for the language `#1'. Using the pattern for}%
\typeout{** the default language instead.}%
\else
\language=\csname l@#1\endcsname
\fi
#2}}
\providecommand{\BIBdecl}{\relax}
\BIBdecl

\bibitem{pochacker2013simulating}
M.~P{\"o}chacker, A.~Sobe, and W.~Elmenreich, ``Simulating the smart grid,'' in
  \emph{IEEE PowerTech (POWERTECH)}.\hskip 1em plus 0.5em minus 0.4em\relax
  IEEE, 2013, pp. 1--6.

\bibitem{muhsen2016multiobjective}
D.~H. Muhsen, A.~B. Ghazali, and T.~Khatib, ``Multiobjective differential
  evolution algorithm-based sizing of a standalone photovoltaic water pumping
  system,'' \emph{Energy Conversion and Management}, vol. 118, pp. 32--43,
  2016.

\bibitem{ramachandran2011intelligent}
B.~Ramachandran, S.~K. Srivastava, C.~S. Edrington, and D.~A. Cartes, ``An
  intelligent auction scheme for smart grid market using a hybrid immune
  algorithm,'' \emph{IEEE Transactions on Industrial Electronics}, vol.~58,
  no.~10, pp. 4603--4612, 2011.

\bibitem{kathib-elmenreich:16}
T.~Khatib and W.~Elmenreich, \emph{Modeling of Photovoltaic Systems Using
  {MATLAB}: Simplified Green Codes}.\hskip 1em plus 0.5em minus 0.4em\relax
  Wiley, 2016, iSBN 978-1-119-11810-7.

\bibitem{albadi2008summary}
M.~H. Albadi and E.~El-Saadany, ``A summary of demand response in electricity
  markets,'' \emph{Electric power systems research}, vol.~78, no.~11, pp.
  1989--1996, 2008.

\bibitem{depuru2011smart}
S.~S. S.~R. Depuru, L.~Wang, and V.~Devabhaktuni, ``Smart meters for power
  grid: Challenges, issues, advantages and status,'' \emph{Renewable and
  sustainable energy reviews}, vol.~15, no.~6, pp. 2736--2742, 2011.

\bibitem{van2014increasing}
M.~van~der Kam and W.~van Sark, ``Increasing self-consumption of photovoltaic
  electricity by storing energy in electric vehicle using smart grid technology
  in the residential sector-a model for simulating different smart grid
  programs.'' in \emph{SMARTGREENS}, 2014, pp. 14--20.

\bibitem{lu2010smartgridlab}
G.~Lu, D.~De, and W.-Z. Song, ``Smartgridlab: A laboratory-based smart grid
  testbed,'' in \emph{Smart Grid Communications (SmartGridComm), 2010 First
  IEEE International Conference on}.\hskip 1em plus 0.5em minus 0.4em\relax
  IEEE, 2010, pp. 143--148.

\bibitem{frincu2014accurate}
M.~Frincu, C.~Chelmis, M.~U. Noor, and V.~Prasanna, ``Accurate and efficient
  selection of the best consumption prediction method in smart grids,'' in
  \emph{Big Data (Big Data), 2014 IEEE International Conference on}.\hskip 1em
  plus 0.5em minus 0.4em\relax IEEE, 2014, pp. 721--729.

\bibitem{mets2014combining}
K.~Mets, J.~A. Ojea, and C.~Develder, ``Combining power and communication
  network simulation for cost-effective smart grid analysis,'' \emph{IEEE
  Communications Surveys \& Tutorials}, vol.~16, no.~3, pp. 1771--1796, 2014.

\bibitem{razaq2015simulating}
A.~Razaq, B.~Pranggono, H.~Tianfield, and H.~Yue, ``Simulating smart grid:
  {Co-simulation} of power and communication network,'' in \emph{Power
  Engineering Conference (UPEC), 2015 50th International Universities}.\hskip
  1em plus 0.5em minus 0.4em\relax IEEE, 2015, pp. 1--6.

\bibitem{bian2015real}
D.~Bian, M.~Kuzlu, M.~Pipattanasomporn, S.~Rahman, and Y.~Wu, ``Real-time
  co-simulation platform using {OPAL-RT} and {OPNET} for analyzing smart grid
  performance,'' in \emph{2015 IEEE Power \& Energy Society General
  Meeting}.\hskip 1em plus 0.5em minus 0.4em\relax IEEE, 2015, pp. 1--5.

\bibitem{hansen2016enabling}
T.~M. Hansen, R.~Kadavil, B.~Palmintier, S.~Suryanarayanan, A.~A. Maciejewski,
  H.~J. Siegel, E.~K. Chong, and E.~Hale, ``Enabling smart grid cosimulation
  studies: Rapid design and development of the technologies and controls,''
  \emph{IEEE Electrification Magazine}, vol.~4, no.~1, pp. 25--32, 2016.

\bibitem{ahourai2013grid}
F.~Ahourai and M.~A. Al~Faruque, ``Grid impact analysis of a residential
  microgrid under various ev penetration rates in gridlab-d,'' \emph{Center for
  Embedded Computer Systems, Irvine, CA}, 2013.

\bibitem{owuor2011ieee}
J.~O. Owuor, J.~L. Munda, and A.~A. Jimoh, ``The ieee 34 node radial test
  feeder as a simulation testbench for distributed generation,'' in
  \emph{AFRICON, 2011}.\hskip 1em plus 0.5em minus 0.4em\relax IEEE, 2011, pp.
  1--6.

\bibitem{moffet2011review}
M.-A. Moffet, F.~Sirois, and D.~Beauvais, ``Review of open-source code power
  grid simulation tools for long-term parametic simulation,''
  \emph{CanmetENERGY, Tech. Rep.}, vol. 137, 2011.

\bibitem{pochacker2014microgrid}
M.~P{\"o}chacker, T.~Khatib, and W.~Elmenreich, ``The microgrid simulation tool
  {RAPSim}: Description and case study,'' in \emph{2014 IEEE Innovative Smart
  Grid Technologies-Asia (ISGT ASIA)}.\hskip 1em plus 0.5em minus 0.4em\relax
  IEEE, 2014, pp. 278--283.

\bibitem{sobe2013smart}
A.~Sobe and W.~Elmenreich, ``Smart microgrids: Overview and outlook,'' in
  \emph{Proceedings of the ITG INFORMATIK Workshop on Smart Grids},
  Braunschweig, Germany, Sep. 2012.

\bibitem{elmenreich2013demand}
W.~Elmenreich and S.~Schuster, ``Demand response by decentralized device
  control based on voltage level,'' in \emph{International Workshop on
  Self-Organizing Systems}.\hskip 1em plus 0.5em minus 0.4em\relax Springer,
  2013, pp. 186--189.

\bibitem{pochacker2015model}
M.~P{\"o}chacker and W.~Elmenreich, ``Model implementation for the extendable
  open source power system simulator {RAPSim},'' in \emph{Intelligent Solutions
  in Embedded Systems (WISES), 2015 12th International Workshop on}.\hskip 1em
  plus 0.5em minus 0.4em\relax IEEE, 2015, pp. 103--108.

\bibitem{schneider2009distribution}
K.~P. Schneider, D.~Chassin, Y.~Chen, and J.~C. Fuller, ``Distribution power
  flow for smart grid technologies,'' in \emph{Power Systems Conference and
  Exposition, 2009. PSCE'09. IEEE/PES}.\hskip 1em plus 0.5em minus 0.4em\relax
  IEEE, 2009, pp. 1--7.

\bibitem{chassin2014gridlab}
D.~P. Chassin, J.~C. Fuller, and N.~Djilali, ``Gridlab-d: An agent-based
  simulation framework for smart grids,'' \emph{Journal of Applied
  Mathematics}, 2014.

\bibitem{tenney2008gridlab}
N.~D. Tenney, ``{GridLAB-D} technical support document: Climate module version
  1.0,'' Pacific Northwest National Laboratory (PNNL), Richland, WA (US), Tech.
  Rep., 2008.

\end{thebibliography}

\end{document}